\documentclass[structabstract]{aa}  
\usepackage{graphicx}
\usepackage{txfonts}
\usepackage{natbib}
\bibpunct{(}{)}{;}{a}{}{,}

\def\kms    {\ifmmode{{\rm ~km\,s}^{-1}}\else{~km\,s$^{-1}$}\fi}

\def\degree   {$\rm ^ o$}

\begin{document}

\title{The AMIGA sample of isolated galaxies XI.}
\subtitle{A First Look at Isolated Galaxy Colors}

\author{M. Fern{\'a}ndez Lorenzo\inst{1}, J. Sulentic\inst{1}, L. Verdes--Montenegro\inst{1}, J. E. Ruiz\inst{1}, J. Sabater\inst{2} \and S{\'a}nchez, S.\inst{1}}
\offprints{M. Fern{\'a}ndez Lorenzo}
\institute{\inst{1}Instituto de Astrof{\'i}sica de Andaluc{\'i}a, Granada,  IAA-CSIC Apdo. 3004, 18080 Granada, Spain \email{mirian@iaa.es} \\
\inst{2}Institute for Astronomy, University of Edinburgh, Edinburgh EH9 3HJ, UK 
}
 \date{Received.....; accepted..... }

\abstract
{The basic properties of galaxies can be affected by both nature (internal processes) or nurture (interactions and effects of environment). Deconvolving the two effects is an important current effort in astrophysics. Observed properties of a sample of isolated galaxies should be largely the result of
internal (natural) evolution. It follows that nurture--induced galaxy evolution can only be understood through comparitive study of galaxies in different environments.}
{We take a first look at SDSS (g--r) colors of galaxies in the AMIGA sample involving many of the most isolated galaxies in the local Universe. This leads us to simultaneously consider the pitfalls of using automated SDSS colors.}
{We focus on median values for the principal morphological subtypes found in the AMIGA sample (E/S0 and Sb--Sc) and compare them with equivalent measures obtained for galaxies in denser environments.}
{We find a weak tendency for AMIGA spiral galaxies to be redder than objects in close pairs. We find no clear difference when we compare with galaxies in other (e.g. group) environments. However, the (g$-$r) color of isolated galaxies shows a Gaussian distribution as might be expected assuming nurture--free evolution. We find a smaller median absolute deviation in colors for isolated galaxies compared to both wide and close pairs. The majority of the deviation on median colors for spiral subtypes is caused by a color--luminosity correlation. Surprisingly isolated and non--isolated early--type galaxies show similar (g$-$r). We see little evidence for a green valley in our sample with most spirals redder than (g$-$r)=0.7 having spurious colors.}
{The redder colors of AMIGA spirals and lower color dispersions for AMIGA subtypes --compared with close pairs-- is likely due to a more passive star formation in very isolated galaxies.}

\keywords{galaxies: general --- galaxies: fundamental parameters --- galaxies: interactions --- galaxies: evolution}

\titlerunning{The AMIGA sample of isolated galaxies XI}
\authorrunning{Fern{\'a}ndez Lorenzo et al.}
\maketitle

\section{Introduction}

In some ways our study of galaxy properties is still in its infancy. We have a plethora of theoretical ideas but we are still trying to understand what basic measures e.g. size, luminosity and color are telling us. This is, in part, because galaxies are composite structures and in another part because we cannot easily separate effects of nature (internal processes) from nurture (interactions and effects of environment).

The optical colors of galaxies reflect their stellar populations and these colors correlate with morphology and environment. The distribution of galaxy colors in the (g$-$r) vs. (u$-$g) plane \citep{2001AJ....122.1861S} shows a strong bimodality with clear separation into red and blue sequences. Study of morphology and spectral classification \citep{2001AJ....122.1861S} for subsamples of 287 red and 500 blue galaxies show that the two color peaks correspond roughly to early-- (E, S0, and Sa) and late--type (Sb, Sc, and Irr) galaxies as expected from the respective dominance of old and young stellar populations. Colors of galaxies also depend strongly on luminosity in the sense that more luminous galaxies of the same morphological type are redder \citep{2004ApJ...600..681B}. The color--magnitude relation is most obvious in the rest--frame (g$-$r) (corrected to z=0.1), where the separation between the red and blue populations is evident \citep{2003ApJ...594..186B}.

Environment is also thought to play a role in the mix of morphological types for a sample of galaxies which is reflected by the morphology--density relation \citep[][and references therein]{1980ApJ...236..351D,2010ApJ...714.1779V}. In dense environments luminous red early--type galaxies predominate while in lowest density environment blue late--type spirals are the defining population \citep{1980ApJ...236..351D,2007ApJS..172..284C}. While it is easy to recognize a rich cluster, definitions of low density environments can be confusing. In recent years there has been an increased emphasis on identifying low density or isolated galaxy populations. One of the most useful samples remains the visually selected Catalog of Isolated Galaxies (CIG) compiled by \citet{1973AISAO...8....3K} more recently vetted as the AMIGA sample \citep[][and references therein]{2010ASPC..421....3S}.

Star formation is strongly dependent on environment with increased activity towards low density environments \citep{2002MNRAS.334..673L, 2003ApJ...585L...5H,2004ApJ...601L..29H, 2006MNRAS.373..469B}. In this sense, \citet{1998ApJ...504L..75B} found that the mean star formation rate in galaxies 
with similar bulge--to--total (B/T) luminosity ratios is always lower in clusters than in the field. However \citet{2011MNRAS.412..591P} found that the opposite is happening in galaxy pairs with clear signs of star formation induced by interaction within the blue galaxies. This result agrees with the bluer (U$-$B and B$-$V) colors found previously by \citet{1978ApJ...219...46L} for peculiar galaxies. 

Approximately 10\% of the low density universe is populated by pairs (also by triplets and dense groups) where environmental effects can reproduce the processes that happen in starburst galaxies and clusters \citep[e.g.][]{1991ApJ...374..407X,2002AJ....124.2471I}. What we are trying to say is that there are two sources of environmental effects: 1) the morphology--density relation and 2) one--on--one galaxy interactions which can be found in a wide range of density environments. In this sense, the AMIGA vetting of the CIG (see below) is intended to produce an isolated sample that minimizes both effects. 

The AMIGA project (Analysis of the interstellar Medium of Isolated GAlaxies) is producing and analyzing a multiwavelength database for a refinement of the Catalog of Isolated Galaxies \citep[CIG,][n=1050 galaxies]{1973AISAO...8....3K}. We have evaluated and improved the sample in different ways: 1) Revision of all CIG positions in the sky \citep{2003A&A...411..391L}, 2) optical characterization of the sample including completeness, luminosities, and heliocentric velocities \citep[][and section 2 of this paper]{2005A&A...436..443V}, 3) Morphological refinement and identification of galaxies with asymmetries \citep[][and section 2.2 of this paper]{2006A&A...449..937S}, and 4) quantification of the degree of isolation including the number density to the 5th neighbour ($\eta_k$) and the estimation of the tidal force (Q) \citep{2007A&A...470..505V,2007A&A...472..121V}.

The multiwavelength analysis of our AMIGA sample has shown that isolated galaxies have different properties than galaxies in higher density environments (even field samples). Variables expected to be enhanced by interactions are lower in AMIGA than any other sample, such as lower infrared luminosity (L$_{FIR}$$<$ 10.5 L$\sun$) and colder dust temperature \citep{2007A&A...462..507L}, a low level of radio continuum emission dominated by mild disk star formation \citep{2008A&A...485..475L}, no radio active galactic nuclei (AGN) selected using the radio--FIR correlation \citep[0$\%$;][]{2008A&A...486...73S} and a small number of optical AGN \citep[22$\%$;][submitted]{Sabater2012}, a smaller fraction of asymmetric HI profiles \citep[$<$ 20$\%$,][]{2011A&A...532A.117E}, and less molecular gas \citep{2011arXiv1108.2130L}. In addition, early--type galaxies in AMIGA are fainter than late types, and more AMIGA spirals host pseudo--bulges rather than classical bulges \citep{2008MNRAS.390..881D}. The data are being released and periodically updated at http://amiga.iaa.es, where a Virtual Observatory compliant web interface with different query modes has been implemented.

This paper presents a first look at colors for the AMIGA sample. It now becomes possible to analyze more than half of the AMIGA sample using uniform digital images, magnitudes and colors of the sloan digital sky survey \citep[SDSS,][]{2000AJ....120.1579Y}. In Sect. 2, the data revision of the AMIGA sample is provided. The sample selection is presented in Sect. 3, and the determination of the absolute magnitudes is described in Sect. 4. The dependence of the rest--frame color as function of the morphological type and environment is analyzed in Sect. 5 and 6, respectively. Finally, the conclusions are presented in Sect. 7.

% \citep{2005A&A...436..443V}, which is composed by the most isolated galaxies that exist in the local universe
\begin{table*}
\caption{Main properties of the CIG sample$^{1}$.}
\label{magni}
\begin{center}
\begin{tabular}{rcccccccccccccc}
\hline
{\bf CIG} & {\bf V$_{hel}$} & {\bf D} & {\bf T} & {\bf IA} & {\bf B$_T$} & {\bf A$_{g}$} & {\bf A$_{i}$} & {\bf A$_{K}$} & {\bf B$_{T}^c$} & {\bf L$_B$} & {\bf $D_{25}$} & {\bf $d_{25}$} & {\bf i} & {\bf Target}\\
    & {\bf  (\kms) }      & {\bf  (Mpc) }   & {\bf (RC3)} &   &  {\bf (mag)}  & {\bf (mag)}  & {\bf (mag)}  &  {\bf (mag)} &   {\bf (mag)}  & {\bf [L$_{\odot}$)]} & {\bf (\arcmin)} & {\bf (\arcmin)} & {\bf (\degree)} & \\
{\bf (1)}  & {\bf (2)}  &  {\bf (3)} & {\bf (4)} & {\bf (5)} & {\bf (6)} & {\bf (7)} & {\bf (8)} & {\bf (9)} & {\bf (10)} & {\bf (11)} & {\bf (12)} &  {\bf (13)}  & {\bf (14)}  & {\bf (15) }\\
\hline\hline
 1 & 7299 & 96.926 & 5 & 1 & 14.167 & 0.173 & 0.571 & 0.040 & 13.383 & 10.570 & 1.393 & 0.640 & 65.094 & 0 \\
 2 & 6983 & 94.745 & 6 & 0 & 15.722 & 0.255 & 0.278 & 0.031 & 15.157 & 9.840 & 0.721 & 0.502 & 46.816 & 1 \\
 3 & - & - & 4 & 0 & 16.057 & 0.246 & 0.354 & - & 15.457 & - & 0.402 & 0.242 & 55.052 & 0 \\
 4 & 2310 & 31.880 & 3 & 0 & 12.818 & 0.252 & 0.863 & 0.017 & 11.685 & 10.283 & 3.289 & 0.729 & 90.000 & 1 \\
 5 & 7865 & 105.907 & 0 & 0 & 15.602 & 0.225 & 0.131 & 0.118 & 15.128 & 9.949 & 0.726 & 0.318 & 75.468 & 1 \\
 6 & 4528 & 61.623 & 7 & 1 & 15.395 & 0.412 & 0.638 & 0.016 & 14.329 & 9.798 & 0.697 & 0.308 & 65.321 & 1 \\
 7 & 12752 & 169.896 & 4 & 0 & 15.662 & 0.100 & 0.338 & 0.083 & 15.141 & 10.354 & 0.692 & 0.427 & 53.958 & 1 \\
 8 & 6342 & 85.041 & 5 & 0 & 15.624 & 0.489 & 0.726 & 0.035 & 14.374 & 10.060 & 0.752 & 0.277 & 71.680 & 0 \\
 9 & 8474 & 113.035 & 5 & 1 & 15.537 & 0.111 & 0.637 & 0.047 & 14.742 & 10.160 & 0.910 & 0.380 & 68.066 & 1 \\
 10 & 4613 & 63.645 & 5 & 0 & 15.377 & 0.366 & 0.546 & 0.025 & 14.440 & 9.782 & 1.054 & 0.501 & 63.935 & 0 \\
.. & .. & ..& .. & .. & .. & .. & .. & .. & .. & .. & .. & .. & .. & ..\\
 \hline
\end{tabular}
 \begin{list}{}{}
 \item[$^{\rm 1}$] The full table is available in electronic form at http://amiga.iaa.es. The columns correspond to
(1) Galaxy identification according to CIG catalog; (2) Heliocentric velocity; (3) Distance;
(4) Morphological type; (5): Degree of optical asymmetry, (6) B--band magnitude from HyperLeda, (7):
Galactic extinction, (8): Internal extinction; (9) K correction in the B--band; (10): Corrected magnitude in the B--band; (11) logarithm of the optical luminosity in the B--band; (12): Major axis; (13) Minor axis; (14) Galaxy
inclination, as described in Sect. 2.  In column (15) we flag galaxies wich code 1 
when the galaxy is part of the studied in this paper, and 0 otherwise, following Sect. 3.
 \end{list}
\end{center}
\end{table*}

\section{The AMIGA data revision}

%CIG & Velocity & Distance & T & m$_{B}$& m$_{B-corr}$& L$_B$        & D$_{25}$ $\times$ d$_{25}$ &  a   & Inclination \\
%       &  (\kms)       &  (Mpc)    & (RC3)    &    (mag)     &        (mag)        & (L$_{\odot}$)& (\arcmin $\times$ \arcmin)       & (kpc) & 
\footnotetext[1]{http://www.wf4ever-project.org}
\footnotetext[2]{http://www.taverna.org.uk/}
\footnotetext[3]{http://www.myexperiment.org/packs/231.html}

Given that, since the project started, a significant number of new data are available for the sample, we have revised the apparent magnitudes, morphological types, distances and optical luminosities for the CIG sample with respect to the ones used in \citet{2005A&A...436..443V} and \citet{2007A&A...462..507L}. In the frame of the Wf4Ever project$^1$, the revision of these properties has been partially automated by the implementation of scientific workflows which ask for and gather values, of some of the properties for all the AMIGA galaxies, from the HyperLeda catalog (see details below). These workflows allow also comparison between values stored in the AMIGA database coming from different releases, registering the new values while keeping a version of the old ones. Therefore the methodology to calculate these values is stored in these workflows enabling reproducibility and re--usability, even the re--purposability for similar cross--boundary use cases. Currently releases of data in HyperLeda are not registered, hence while the methodology, as explained above, can be preserved, traceability of the HyperLeda data is not possible yet. These workflows have been built in the Taverna Workflow Mamagement System$^2$ and can be accessed in the MyExperiment portal$^3$. The data for the full AMIGA sample are listed in Table~\ref{magni}, and are also available in the AMIGA VO compliant interface. We describe the details of the revision below. For completeness, we explain here all the revised parameters, although optical luminosity has not been used in this paper.

\subsection{Velocities and Distances \label{subsec:velocities}}

We have searched for new velocities for those CIG with heliocentric velocities (V$_{hel}$) larger than 1000 km/s, since for lower values redshift--independent distance estimates have been preferred, and the same as in \citet{2005A&A...436..443V} were kept. The value of V$_{hel}$ for a galaxy with available redshift data has been updated when the errors quoted in the recent bibliography were smaller than in our previous compilation. For a subset of galaxies we found that the source of the data provided in NED were obtained from \citet{2004AJ....127.1336S} but, while the paper was providing galactocentric velocities, NED was quoting them as heliocentric. This error was reported to NED team and both corrected in NED and in our database. 
Only for 40 CIG galaxies velocity data were not found in the bibliography, hence 1010 out of 1050 CIG galaxies currently have redshift measures.
%In Fig.~\ref{fig:VELOCITY} we compare the velocity distribution for all galaxies in the CIG sample with known redshift and those galaxies with \hi\ data.

Distances were obtained from virgo--centric velocities (V$_{vir}$) as D = V$_{vir}$/H$_0$ (using H$_0$ = 75 km/s Mpc$^{-1}$), where V$_{vir}$ is calculated from the heliocentric velocity V$_{hel}$ and galactic coordinates following HyperLeda convention. Virgo--centric velocities were calculated as V$_{hel}$ = V$_{lg}$ + 208 cos($\theta$), where V$_{lg}$ is the radial velocity with respect to the Local Group and $\theta$ is the angular distance between the galaxy and Virgo center. In order to transform the V$_{hel}$  to V$_{vir}$ we transformed $\alpha$ and $\delta$ coordinates in \citet{2003A&A...411..391L} to $l$ and $b$ values.

\subsection{Morphologies \label{subsec:MORPHOLOGY_HG84}}

\citet{2006A&A...449..937S} performed a careful revision of the morphologies for the whole CIG sample based on POSS II \citep[Second Palomar Observatory Sky Survey;][]{1991PASP..103..661R} images. We have recently revised those morphologies for all CIG galaxies with V$_{hel}$ $>$ 1000 km/s and CCD images available either from SDSS or our own data (N = 843). For CIG galaxies with V$_{hel}$ $<$ 1000 km/s (N = 57), morphological types were compiled from the bibliography and the mean value used. For N = 134 galaxies with only POSS II data available, a second revision has been performed, with minor modifications to the values assigned in \citet{2006A&A...449..937S}. For the remaining galaxies for which we were unable to perform any classification, we used data found in NED and HyperLeda database.

Globally there is a shift in the morphologies with respect to \citet{2006A&A...449..937S} towards later types by $\Delta$T = 0.2, that we interpret as due to the higher resolution provided by CCD images revealing smaller bulges. Types (T) have been coded following RC3 (see morphological codes in Tab.~\ref{table:1}). \citet{2006A&A...449..937S} also flagged galaxies suspected to be interacting. We have now replaced this code by a descriptive one, based also on visual inspection of the optical images, as described next. Code IA=0 is assigned if no relevant signs of distortions are visible. Code IA=1 corresponds to galaxies seen in the images as asymmetric, lopsided, warped or distorted, with an integral sign shape or tidal feature (tail, bridge, shell). Those galaxies are always treated separately in our analysis, in order to check whether they present a different
behaviour than the ones classified with code 0. Code IA=2 is assigned if the galaxy was identified as pairs in \citet{2007A&A...470..505V} and/or in NED, or if the galaxy looks like a merger or superposition of two galaxies. These galaxies are not considered part of AMIGA sample and are marked differently in the plots. The new data are presented in Tab.~\ref{magni} and now all CIG galaxies have a morphological code. 

\subsubsection{Apparent magnitudes}

Blue optical magnitudes  B$_{T}$ have been obtained from HyperLeda \citep{2003A&A...412...57P}, instead of using the Catalog of Galaxies and Clusters of Galaxies \citep{Zwicky1968} as in \citet{2005A&A...436..443V}, and the corrected B$_{T}^c$ was calculated as follows:

\begin{equation} 
{\rm B_{T}^c = B_{T}-A_{g}-A_{i}-A_{K}}
\end{equation}
where A$_{g}$  is the galactic dust extinction, A$_{i}$ the internal extinction correction and A$_{K}$ the K correction.
A$_{g}$ has been  obtained from HyperLeda \citep{1998ApJ...500..525S}, except for CIG 447 for which only NED data were available.
A$_{i}$ was calculated as in HyperLeda \citep{1995A&A...296...64B} using their values for log(R$_{25}$) (axis ratio of the isophote 25 mag/arcsec2 in the B--band for galaxies) but our own morphological types. Finally A$_{K}$ was obtained as defined in HyperLeda$^2$ \citep[A$_{K}$=ak(T)V$_{hel}$/10000;][]{Vaucouleurs}, using our own morphologies and distances.
\footnotetext[2]{http://leda.univ-lyon1.fr/leda/param/btc.html}
Both uncorrected and corrected magnitudes are listed in Tab.~\ref{magni}.

%{\color{red} -Discussion about different systems for mb, etc. here or as  as an appendix?}

\subsection{Optical luminosity}

The optical luminosities have been derived as

\begin{equation}
{\rm log(L_B/L_{\odot}) = 11.95 + 2 log[D(Mpc)] - 0.4 B_{T}^c}
\end{equation}
where the solar luminosity is given in units of the solar bolometric luminosity as in \citet{2007A&A...462..507L}.
The mean value of the difference between our new values for optical luminosities with respect to the ones used in \citet{2007A&A...462..507L} is as small as 0.02 $\pm$ 0.18 mag but has the advantage of being derived from $B_{T}$ magnitudes compiled from HyperLeda database, so reducing the errors. %The distribution of $L_{B}$ for the \hi\ sample is presented in Fig.~\ref{fig:LB_HI_CIG-all}. 

\subsection{Optical diameter and axis ratio (D$_{25}$ and R$_{25}$)} 

We have made use of the HyperLeda database in the compilation of the major axis D$_{25}$ (isophotal level at 25 mag/arcsec$^2$ in the B--band) and axis ratios $R_{25}$ ($R_{25}$ = $D_{25}$/$d_{25}$, where $d_{25}$ is the minor axis, at an isophotal level of 25 mag/arcsec$^2$), as well as their errors (minor and major axis are given in Tab.~\ref{magni}). 

\subsection{Inclinations \label{subsubsec:inclinations}}

The inclination ($i$) was determined from the value of R$_{25}$ in HyperLeda database and our revised morphological type with the following equation \citep{1972MmRAS..75..105H}:

\begin{equation}
{\rm sin^{2}(i)=\frac{1-10^{-2 log(R_{25})}}{1-10^{-2log(R_{0})}}}
\label{eq:inclination}
\end{equation}

\noindent where log($R_{0}$) = 0.43 + 0.053 $\times$ T  for T$\le$ 7  and log($R_{0}$) = 0.38 for T $>$ 7.

\section{Sample selection}

As starting sample we use the catalog of 791 AMIGA galaxies selected by \citet{2007A&A...472..121V}, where the galaxies with isolation parameters Q$>$-2 and $\eta_k$$>$2.4 \citep[the tidal strength created by all the neighbours, $Q$, is more than 1$\%$ of the internal binding forces; for the local number density, $\eta_k$, this translates to a value of 2.4;][]{atan1984}, and with recession velocities V$_r$$<$1500 km/s, were rejected. These conditions imply that all galaxies in our starting sample have their evolution dominated by their intrinsic properties \citep[see][for more details]{2007A&A...472..121V}. This sample is complete up to B$_{T}^c$=15.3, where B$_{T}^c$ is the magnitude in the B--band from HyperLeda after corrections. The completeness limit was decided in \citet{2005A&A...436..443V} (B$_{T}^c$=15 mag) after applying the $<$V/Vm$>$ test \citep{1968ApJ...151..393S}, and has been recently updated since B--band magnitudes were recalculated following HyperLeda expressions in 2010 (see section 2). There are 657 objects in the complete AMIGA sample that also fulfill the above isolation criteria. Photometric data used in this work come from the SDSS--III \citep[Data Release 8, DR8][]{2011ApJS..193...29A}. The SDSS project used a 2.5 m telescope \citep{2006AJ....131.2332G} to obtain photometric information from CCD images in u, g, r, i, and z passbands. A new approach for background subtraction was applied in DR8 that first models the brightest galaxies in each field in order not to affect the estimated sky background \citep{2011arXiv1105.1960B}. 

 \begin{figure}[t]
\centering
      \includegraphics[angle=0,width=9.0cm]{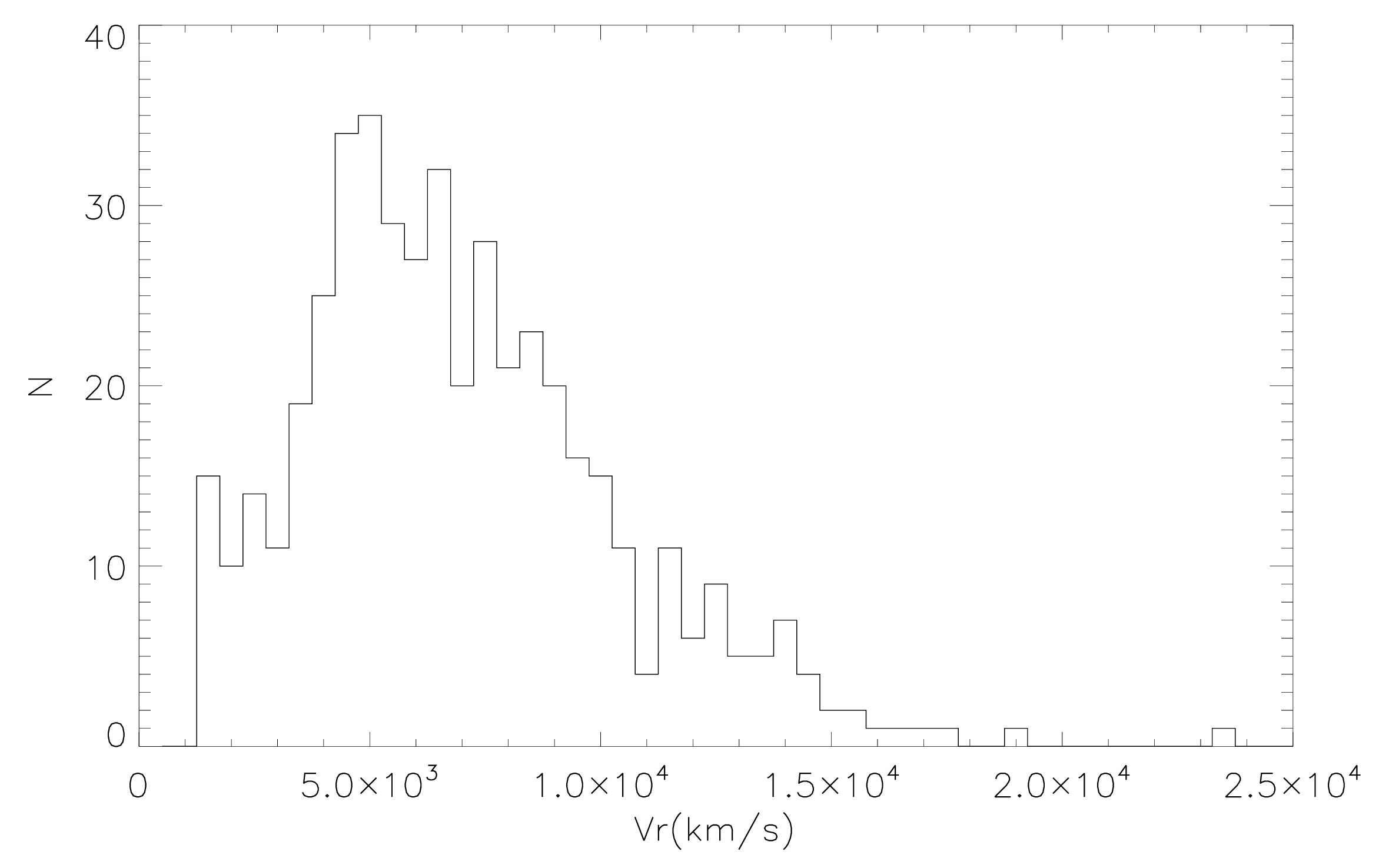}
      \caption{Distribution of the heliocentric velocities of the 466 CIG galaxies used in this work.}
\label{hv}
   \end{figure}

From the 657 galaxies that are part of the complete AMIGA sample, we found 496 objects in the SDSS database. We found and removed 14 galaxies from the catalog involving imprecise identifications of galaxy nuclei in the SDSS database or galaxies with a nearby bright star adversely affecting galaxy photometry. We also removed 16 galaxies with unknown redshifts since those will be needed for the following analysis. The final sample consists of 466 isolated galaxies with radial velocities between 1500 and 24000 $\rm km \ s^{-1}$. Hereafter, we refer to it as the AMIGA--SDSS sample. We have represented its redshift distribution in Fig.~\ref{hv}. This sample includes 70 galaxies with asymmetries that are suspected of being involved in interactions or having nearby companions. These properties are represented in our sample by the degree of asymmetry IA=1 (63) or IA=2 (7) (see Section 2.2), where IA=2 represents the most asymmetric galaxies. We do not reject these objects since we are trying to see what features can a galaxy develop in isolation, but, throughout the paper, we check the effect of these galaxies in the median colors. We focus our attention on the two most statistically significant subsamples involving Sb--Sc spirals and E/S0 early--types. These involve approximately 68\% and 15\% of the AMIGA--SDSS sample used in this work respectively. The former is interpreted as a kind of "parent population" of isolated galaxies because it represents 2/3 of the total sample. The latter was somewhat of a surprise under the assumption that isolated galaxies are simply the outliers of loose groups where the early--type fraction is $\sim$ 0.4 \citep{2007ApJ...670..206V}. We find $\sim$0.15 in our sample, which is still large enough to be interesting. Other types are represented in such small numbers that we assume them to be simply random galaxies that happen to be unusually isolated at this time. This assumption is motivated by the standard idea that all disky galaxies with a large bulge are products of nurture. The late--type spiral population involves galaxies with small bulges \citep[or even pseudo--bulges,][]{2008MNRAS.390..881D} which, if interpreted as a secular evolution diagnostic, cannot have spent much time in richer environments. 

\section{Deriving rest--frame colors}

A variety of magnitude estimates are given for each galaxy included in the SDSS catalog.
We chose the model magnitudes for deriving colors in our sample because they are calculated using best--fit parameters in r--band which are then applied to the other bands. Model magnitudes are therefore computed through the same aperture for all bands. In order to compute absolute magnitudes in each band the following corrections were applied:

\begin{itemize}
\item A correction for Galactic dust extinction applying the reddening corrections computed by SDSS following \citet{1998ApJ...500..525S}.

\item The k--correction in each band was calculated using the code {\tt kcorrect} \citep{2007AJ....133..734B}, which determines the spectral energy distribution (SED) of the galaxy using SDSS photometry from a nonnegative linear combination of five templates based on the \citet{2003MNRAS.344.1000B} stellar evolution synthesis codes.

\item Absolute magnitudes were calculated using the updated distances shown in Tab.~\ref{magni}.

\end{itemize}

The internal extinction correction in B--band was calculated by \citet{2005A&A...436..443V} as a function of inclination and morphological type following \citet{1991trcb.book.....D}. We used the Calzetti law \citep{2000ApJ...533..682C} for deriving the SDSS--band internal extinctions based on the B--band corrections. We represented the rest--frame color (g$-$r) versus inclination before and after the extinction corrections and found that this correction was overestimated (the slope of the linear regression changes from 0.17 to $-$0.21 for Sc galaxies and from 0.16 to $-$0.13 for Sb galaxies, after corrections). \citet{2010MNRAS.404..792M} studied the effect of this correction on the colors through its dependency on inclination, spiral type and absolute magnitude. The best result for our (g$-$r) versus inclination plots is obtained using a linear combination of their relations as defined in equation 3 in their paper (the slope of the (g$-$r) versus inclination relation after this correction is 0.025 for Sc and 0.023 for Sb galaxies), while the expressions they give for each morphological type (bulge--disk ratio) produce an overcorrection in all our subsamples. We also note that the dependence of the internal extinction correction on galaxy luminosity derived in \citet{2010MNRAS.404..792M} is not reproduced by our data. If AMIGA spirals are really more dynamically quiescent systems, it is perhaps not surprising that corrections derived from almost certainly less quiescent samples would yield different corrections. Moreover, this preliminary result of lower extinction in AMIGA spirals can be understood as a sign of less dust due to our assumption that the star formation has been lower in our isolated galaxies for all or most of their lives \citep{2008A&A...485..475L}.

Finally, we applied the correction of \citet{2010MNRAS.404..792M}, based on their equation 3, to our Sa--Sc galaxies. We have not corrected the morphological types earlier than 0 (E/S0) and later than 6 (Scd/Sd/Irr) because their numbers are small and their (g$-$r) vs. inclination plots do not show any clear trend of color vs. i (the slope of the fits is $\sim$ 0). 

 \begin{figure}[t]
\centering
      \includegraphics[angle=0,width=9.0cm]{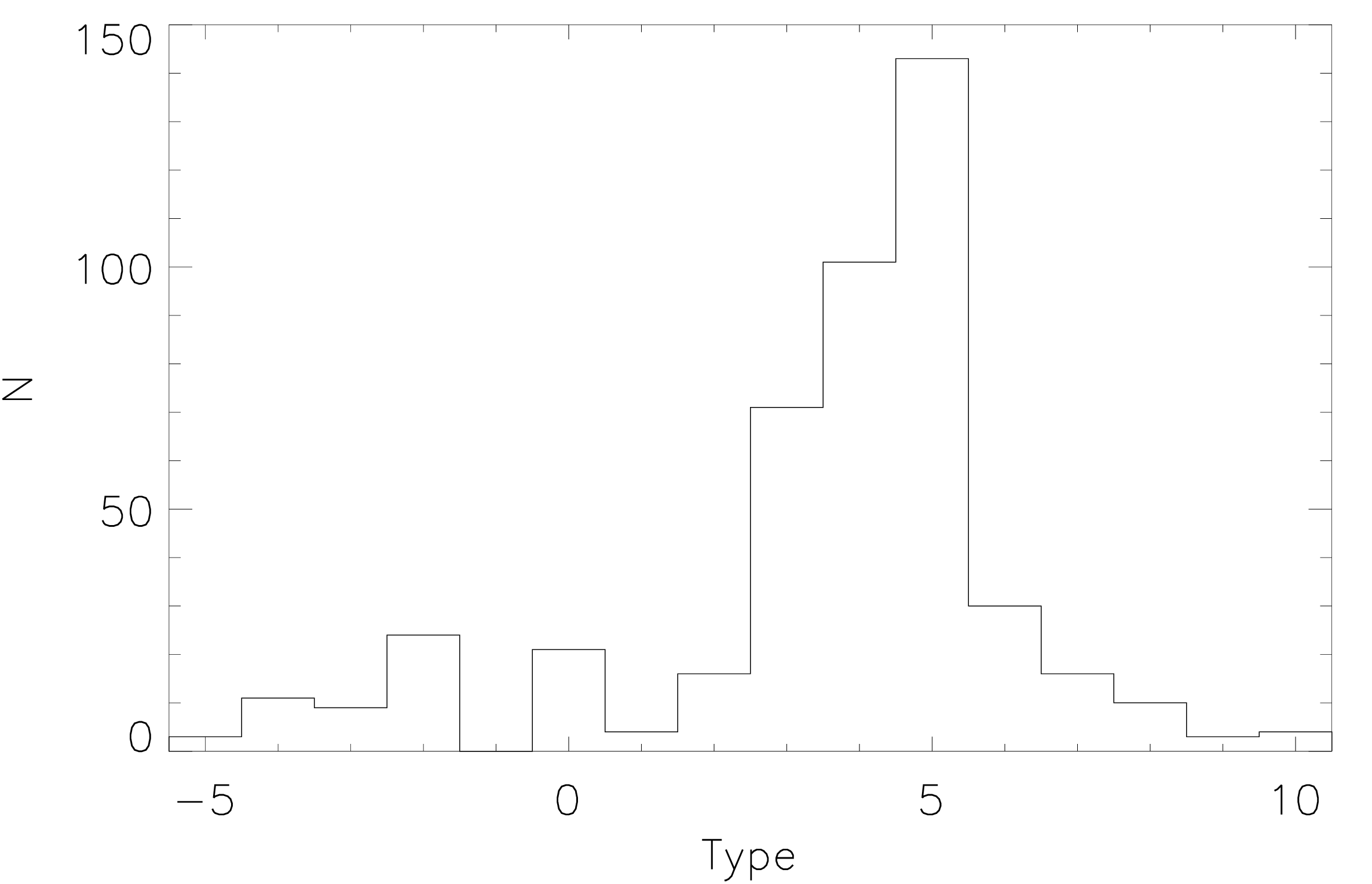}
      \caption{Distribution of Hubble types in the AMIGA--SDSS sample.}
 \label{hm}
   \end{figure}

 \begin{figure}[t]
\centering
      \includegraphics[angle=0,width=9.0cm]{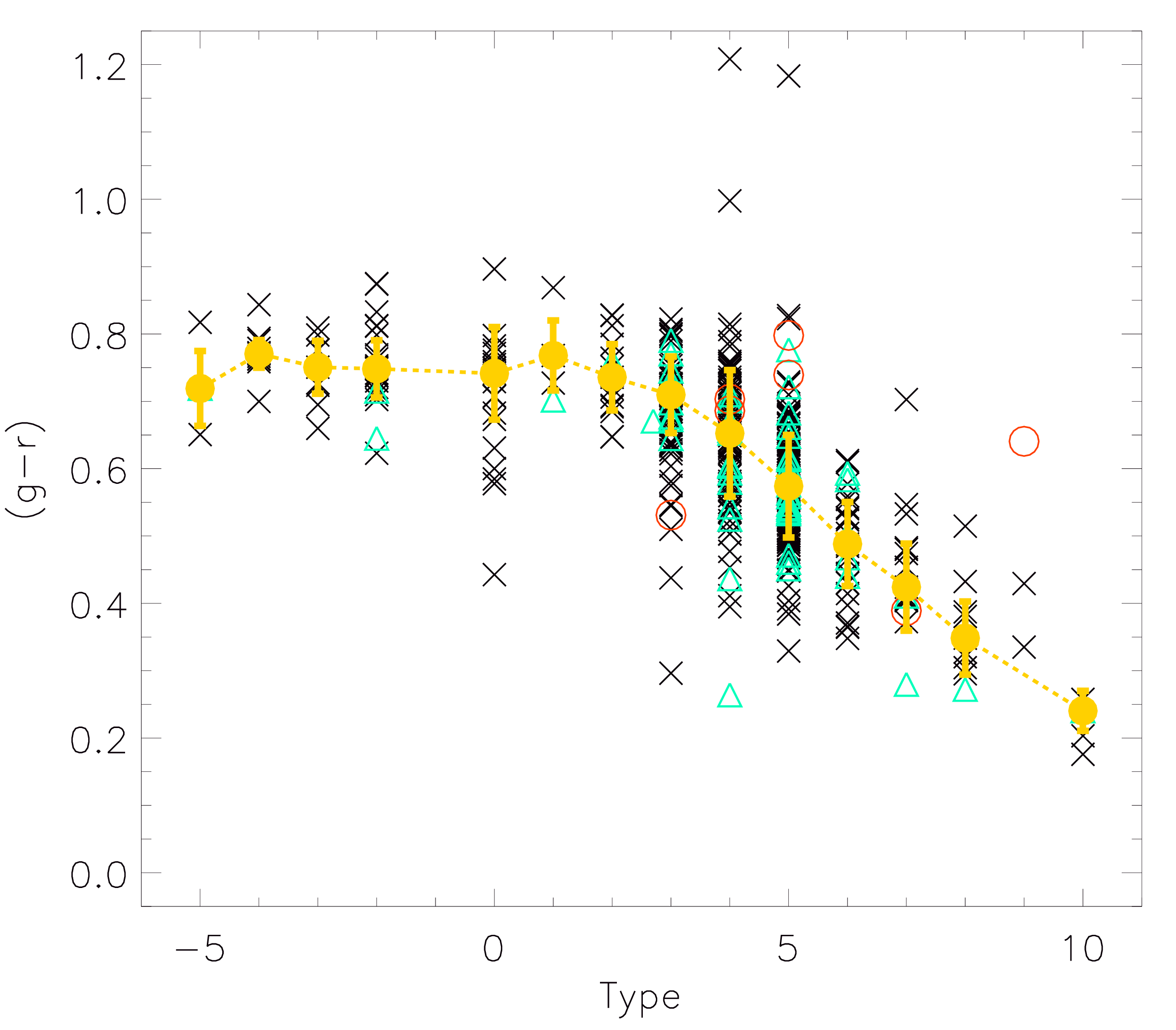}
      \caption{Distribution of the rest--frame (g$-$r) color as a function of Hubble type. The yellow points are the median values of each morphological type. The blue triangles are objects with asymmetry index IA=1 while the red open points represent the most asymmetric objects (IA=2) in our sample. The error bars represent the median absolute deviation.}
 \label{hc}
   \end{figure}

\section{(g--r) color as function of morphological type}

In Fig.~\ref{hm} we presented the number of galaxies in the AMIGA--SDSS sample of each morphological subtype, and Fig.~\ref{hc} shows the distribution of SDSS (g$-$r) colors for the AMIGA sample as a function of the morphological subtype. We calculated the median values rather than mean values, since they are less sensitive to outliers produced by an error in the color for one object or misclassification. We have used the median absolute deviation as a scatter measure since it represents a robust estimator of dispersion for median values. Not surprisingly the reddest median values of (g$-$r) are found for the first four bins representing early--type galaxies although median (g$-$r) values remain essentially constant out to T=3 (Sb). Beginning with type Sb we see a decrease in median (g$-$r) as expected if this sequence reflects a uniformly decreasing contribution from an old stellar population \citep{2001AJ....122.1861S, 2003ApJ...594..186B}.

Surprisingly the color distributions for our late--type spiral Sb--Sc population show some galaxies with colors even redder than the E/S0 subsample. Either AMIGA early--type galaxies are extraordinarily blue or these spirals have anomalous colors. Careful examination of these objects using the SDSS navigation tool shows that colors of the reddest Sb--Sc galaxies are unreliable and/or that the assigned galaxy types are incorrect. In the case of our Sc subsample (which is perhaps the easiest to classify) we find n=10 galaxies with (g$-$r)$>$0.7  (tipical colors of early--types). In 8 cases there is a problem with the color: 1) five are affected by a star projected on or very close to the galaxy (CIG 261, 304, 649, 716, 1010), 2) two (CIG 349, 946) belong to the 7 galaxies previously identified with a high degree of asymmetry (IA = 2), 3) one (CIG 988) is highly inclined and the extinction correction \citep{2010MNRAS.404..792M} could be maximally affected by uncertainty of the adopted inclination. Of the remaining 2 red Sc spirals, one (CIG 709) shows among the highest recession velocity in our sample (V$>$14000km/s). At this distance the number of resolution elements in the galaxy images approaches that of POSS II. In the case of CIG 709, the image of SDSS shows a close object with unknown redshift which could be a minor companion causing some kind of asymmetry in the optical bands. Finally, CIG 807 may well be an Sc spiral but the blue disk is not resolved. The point of this exercise is to show that one can identify remarkably pure isolated (local) spiral subsamples but only if the above pitfalls are taken into account. If these effects are not taken into account the average properties of the isolated subsamples can be blurred beyond statistical utility. 

Consideration of Sbc and Sb spirals reveal similar effects among the reddest galaxies that only detailed surface photometry could correct. Naturally similar problems can afflict the bluest galaxies in each spiral subtype. Colors for 2/4 of the bluest Sc spirals ((g$-$r)$<$0.4) are also likely affected by their nearby bright stars. These effects can be a source of serious scatter in color distributions for any galaxy sample and illustrate the caution needed in using SDSS colors. On the other hand, while the objects with IA=2 are outside the normal trend of the median values, the color of the AMIGA galaxies with asymmetry index IA=1 agrees with the color of symmetric galaxies (IA=0). In the following analysis, objects with IA=2 have been removed but not objects with spurious colors, since they also exist in the samples that we have used as comparison. Nevertheless, we took account of their effects through the error measures. 

We expect to find a smaller color dispersion for spiral subtypes in the AMIGA sample because these galaxies are minimally affected by environmental effects which apparently induce a larger color dispersion. This has been known since the first statistical study of colors for galaxies in interacting pairs and multiplets \citep[i.e. they evolve more passively, e.g.][]{1978ApJ...219...46L}. One of the goals of the AMIGA project is to provide a sample that could better characterize intrinsic galaxy properties and their dispersions.
 
What is the main source of color dispersion for spiral subtypes if we have
achieved our goal of minimizing effects of environmental nurture from our sample?  
Spirals contain a red bulge (redness depending on the nature of the bulge) and 
a blue disk making their global (g$-$r) color sensitive to the aperture used 
in estimating the g and r magnitudes. We checked the (g$-$r) color versus absolute magnitude relation for Sc galaxies in three separate redshift ranges. The results are presented in Fig. 3. We find a tendency for galaxies of fixed absolute magnitude to be (g$-$r)$\sim$0.08 redder at lower recession velocities implying that color measures in our sample could be increasingly affected by bulge light at lower recession velocities. This is a minor but systematic effect. However, an aperture effect will also understimate the absolute magnitude of the galaxy, making it more difficult to estimate the amplitude of the color bias. 

\begin{figure}[t]
\centering
      \includegraphics[angle=0,width=9.0cm]{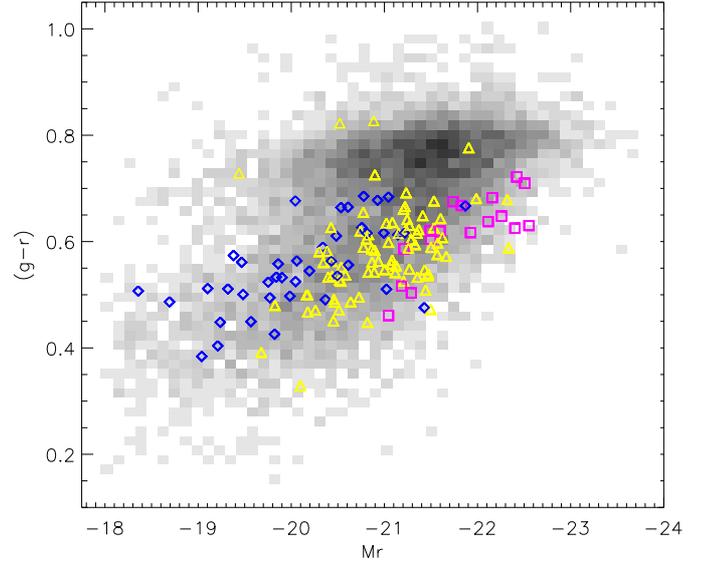}
      \caption{(g$-$r) color--magnitude diagram for the Sc galaxies in the AMIGA sample. The blue diamonds are objects at z$<$0.02, the yellow triangles are galaxies at redshift 0.02$<$z$<$0.04, and the pink squares are those objects at z$>$0.04. The grey--scale represents the density diagram obtained from the \citet{2010ApJS..186..427N} sample at 0.01$<$z$<$0.05, using the data of DR8.}
\label{cm_sc}
   \end{figure}

Fig.~\ref{cm_sc} shows that the major source of color dispersion in Fig.~\ref{hc} is connected with the color--luminosity trend. Sb--Sc color distributions show Gaussian distributions for each subtype with FWHM=0.15--0.2, spanning almost 2dex in luminosity (M$_r$=$-$19 to $-$22.5). The effect is largest for Sc spirals in our sample where the color--luminosity trend appears to be steeper (Sc spirals with M$_r$=-22.5 are $\sim$0.25mag redder than Sc spirals with M$_r$=-19). Fig.~\ref{cm_sc} also shows the (g$-$r) vs. M$_r$ density diagram obtained from the \citet{2010ApJS..186..427N} sample covering the range 0.01$<$z$<$0.05 (SDSS DR8). This sample applied no morphology or environmental selections so we see the bimodality previously found for SDSS data \citep{2001AJ....122.1861S, 2003ApJ...594..186B}. The distribution of our Sc subsample follows the blue sequence with the recession velocity vs. color bias visible. The same bias in the color as function of redshift has been found by using the sample of \citet{2010ApJS..186..427N}. Moreover, for a given luminosity and morphological type (Sc), the petrosian radius R$_{90}$ given in the SDSS database for the \citet{2010ApJS..186..427N} sample decreases as we go at lower redshifts. The main trend of our data in the figure (and hence source of color dispersion) clearly involves the color--luminosity correlation. The overlap (i.e. green valley) rather than being a class of galaxies with intermediate colors may be entirely due to spirals with spurious red colors due to the effects discussed above. The (g$-$r) vs. M$_r$ diagram for a vetted sample of isolated galaxies likely shows a much clearer dichotomy than other samples. There is no evidence for an astrophysically significant green valley but we need more reliable colors to confirm this result.

\begin{figure}[t]
\centering
\includegraphics[angle=0,width=9.0cm]{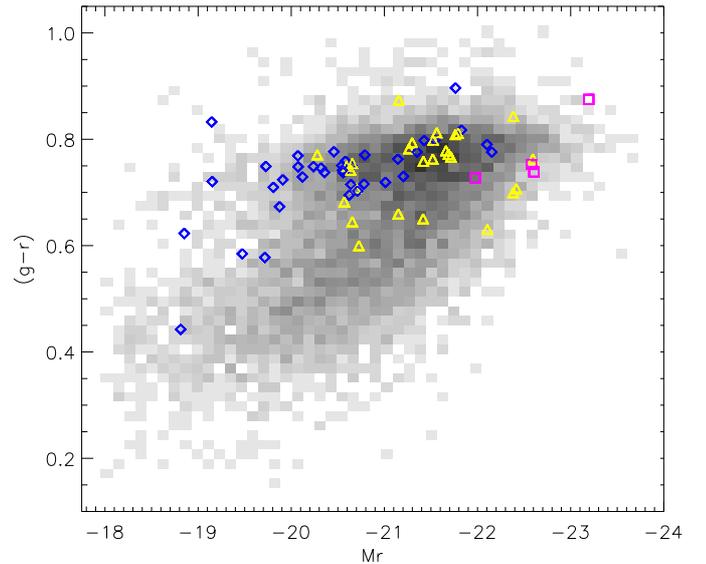}
\caption{Same as Fig.~\ref{cm_sc} but for the E/S0 galaxies in the AMIGA sample.}
\label{cme}
\end{figure}

Fig.~\ref{cme} shows an equivalent diagram for the early--type (E/S0) part of the AMIGA sample. Despite their isolation they fall in the red sequence defined by early--types in richer environments (e.g. \citet{2010ApJS..186..427N}). They do not provide evidence for a green valley. Our early--type subsample shows no color trend with recession velocity. We notice only a slight trend between color and luminosity (galaxies with M$_r$=-22.5 are $\sim$0.1mag redder than galaxies with M$_r$=-19). In the case of early--types the FWHM of the Gaussian fit to the color distribution is $\sim$0.14 (similar to Sb galaxies color distribution). The bluest galaxies might be misclassified as E/S0 although photometric study of a few of them \citep{2004AJ....127.3213M} suggests that they may not be spirals. In addition, three of them (CIG 264, 981, and 1025) having HI emission were revised in order to check for the presence of peculiarities in the morphologies, such as dust lanes, optical shells, blue cores or star formation. All of them were confirmed as early--type galaxies \citep{TesisDani}.

\section{Color dependence on environment}
 
Some studies have shown that galaxies of a fixed morphology in higher density environments are redder \citep{2008MNRAS.383..907B, 2009MNRAS.399..966S} and the reason might be that dense environments suppress star formation \citep{1998ApJ...504L..75B}. On the other hand isolated galaxies are likely to show rather passive star formation leading one to expect spirals in richer environments (e.g. pairs, even the isolated pairs usually included in field galaxies samples) to show more active star formation and hence bluer colors \citep{2011MNRAS.412..591P}. \citet{1978ApJ...219...46L} found larger color dispersion in an extreme nurture sample, whose evolution is completely dominated by external effects. The key word here is dispersion: colors of galaxies in pairs and groups are not systematically bluer but they show a larger color dispersion. 
The AMIGA sample of galaxies involves some of the most isolated objects in the local Universe and are minimally affected by external processes, at least during the past few Gyr. We therefore expect that AMIGA galaxies of a given morphological type will show minimal color dispersion and the above results support that expectation.

 \begin{table*}%[h]
  \caption{Median (g$-$r) colors as function of morphological type. The errors represent the median absolute deviation.}   
%            % title of Table
  \small
\label{table:1}      % is used to refer this table in the text
 \centering                          % used for centering table
\begin{tabular}{c c c c c c c}
\hline\hline
{\bf Type}&{\bf T}&{\bf AMIGA}&{\bf NAIR}&{\bf EFIGI}&{\bf CPG}&{\bf CPG}\\
{}&{}&{}&{}&{}&{\bf (WID)}&{\bf (CLO)}
\\\hline
E & -5 & 0.72$\pm$0.06 & 0.78$\pm$0.03 & 0.78$\pm$0.03 & 0.79$\pm$0.03 & 0.76$\pm$0.03\\
E & -4 & 0.77$\pm$0.02 & $-$ & 0.78$\pm$0.02 & 0.80$\pm$0.08 & 0.79$\pm$0.04\\
E/S0 & -3 & 0.75$\pm$0.04 & 0.76$\pm$0.05 & 0.77$\pm$0.04 & 0.79$\pm$0.06 & 0.77$\pm$0.07\\
S0 & -2 & 0.75$\pm$0.04 & 0.76$\pm$0.04 & 0.76$\pm$0.04 & 0.78$\pm$0.06 & 0.77$\pm$0.06\\
S0 & -1 & $-$ & $-$ & 0.78$\pm$0.06 & 0.72$\pm$0.05 & 0.73$\pm$0.09\\
S0/a & 0  & 0.74$\pm$0.07 & $-$ & 0.76$\pm$0.07 & 0.77$\pm$0.05 & 0.78$\pm$0.04\\
Sa & 1  & 0.77$\pm$0.05 & 0.71$\pm$0.06 & 0.73$\pm$0.05 & 0.72$\pm$0.11 & 0.71$\pm$0.09\\
Sab & 2  & 0.74$\pm$0.05 & 0.69$\pm$0.07 & 0.72$\pm$0.07 & 0.71$\pm$0.10 & 0.67$\pm$0.15\\
Sb & 3  & 0.71$\pm$0.06 & 0.67$\pm$0.08 & 0.71$\pm$0.08 & 0.71$\pm$0.13 & 0.69$\pm$0.12\\
Sbc & 4  & 0.65$\pm$0.09 & 0.61$\pm$0.08 & 0.66$\pm$0.07 & 0.63$\pm$0.12 & 0.59$\pm$0.14\\
Sc & 5  & 0.57$\pm$0.08 & 0.56$\pm$0.08 & 0.62$\pm$0.09 & 0.69$\pm$0.12 & 0.51$\pm$0.15\\
Scd & 6  & 0.49$\pm$0.06 & 0.46$\pm$0.07 & 0.58$\pm$0.09 & 0.55$\pm$0.11 & 0.51$\pm$0.17\\
Sd & 7  & 0.42$\pm$0.06 & 0.42$\pm$0.06 & 0.47$\pm$0.08 & 0.34$\pm$0.17 & 0.43$\pm$0.10\\
Sdm & 8  & 0.35$\pm$0.05 & 0.41$\pm$0.07 & 0.44$\pm$0.12 & 0.48$\pm$0.07 & 0.30$\pm$0.12\\
Sm & 9  & $-$ & 0.36$\pm$0.09 & 0.40$\pm$0.16 & $-$ & 0.56$\pm$0.13\\
Im & 10 & 0.24$\pm$0.03 & 0.33$\pm$0.10 & 0.29$\pm$0.09 & $-$ & 0.29$\pm$0.12\\
\hline
 \end{tabular}
  \end{table*}

Table \ref{table:1} provides a quantitative comparison of (g$-$r) median colors in our sample and in samples involving denser environments. We used three catalogs for this comparison: 1) \citet{2010ApJS..186..427N} which includes a detailed visual classifications for 14,034 galaxies in the SDSS DR4. We selected only objects with morphological classification available and redshift 0.01$<$z$<$0.05 (8976) to better match our AMIGA--SDSS sample (~98$\%$); 2) the EFIGI catalog \citep{2011arXiv1103.5734B} which provides detailed morphological information for a sample of 4458 PGC galaxies also using SDSS DR4. Both samples include galaxies in a wide range of environments; 3) We also compare with the catalog of isolated pairs of galaxies \citep[CPG,][]{1972SoSAO...7....1K}, composed by 1206 objects. It was visually compiled, like CIG, using POSS II and applying an isolation criterion.
 
\begin{figure}[t]
\centering
      \includegraphics[angle=0,width=9.0cm]{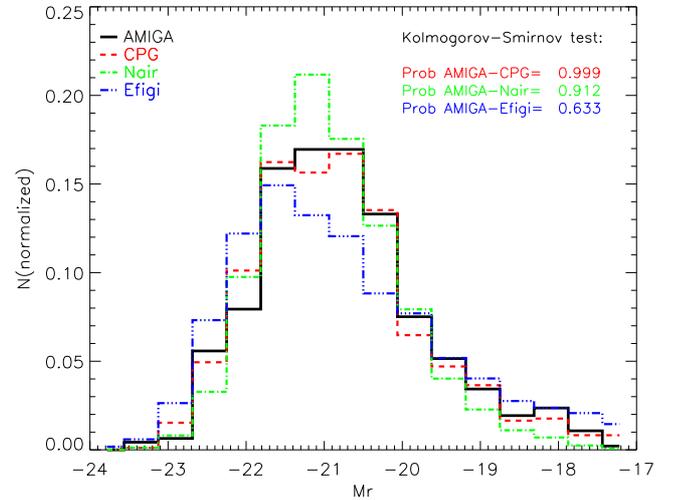}
      \caption{Distribution of absolute magnitudes in the r--band for the whole AMIGA--SDSS sample, and for the \citet{2010ApJS..186..427N}, Efigi, and CPG samples described in the text.}
\label{mags}
   \end{figure}

We used SDSS DR8 photometric data for all three samples and derived rest--frame colors in the same way as for our sample. For the CPG sample, there are 916 galaxies with SDSS photometry. We calculated the projected distance between galaxies in each pair using redshifts taken from HyperLeda. Following \citet{1991ApJ...374..407X} we separated this sample into close--interacting (CLO -- galaxies with separation SEP$<$ 2Mpc) and wide (WID) pairs (SEP$>$2Mpc). We checked the distribution of redshift and absolute magnitudes in the r--band for each sample and found good agreement with the whole AMIGA--SDSS sample. In the cases of EFIGI and CPG catalogs, we removed the objects with recession velocities lower than 1500 km/s as we did for our AMIGA sample, due to the determination of the isolation parameters \citep{2007A&A...472..121V}. We have also removed objects fainter than M$_r>$-17 since we do not have counterparts in the AMIGA--SDSS sample. The final samples of comparison have 8879 (Nair), 3294 (EFIGI), and 839 (CPG) galaxies. In Fig.~\ref{mags} we present the absolute magnitude distribution in the r--band for each sample, as well as the probability given by the Kolmogorov--Smirnov two sample test that these distributions are indistinguishable to the AMIGA--SDSS one. This probability is larger than 90$\%$ for the CPG and Nair samples, while the lowest value is found in the case of the EFIGI sample (63$\%$). 

We found good agreement for the colors of early--type galaxies in all samples. However, in all cases (from T=-5 to T=0), the median values for the AMIGA sample are a little bluer (but within the errors) than the other samples. Because the color--luminosity relation has a trend (brighter galaxies are redder), this bluer color could be presumably due to the fact that the E/S0 population of AMIGA is fainter than any other sample \citep{2006A&A...449..937S}. The median values of colors for the \citet{2010ApJS..186..427N} and close pairs samples, for the (Sb--Sc) spirals, are consistent but slightly bluer than our sample. In the case of the comparison with \citet{2010ApJS..186..427N}, the differences in color may be produced by the different  morphological classifications used in both samples. We find their morphologies to be earlier than ours, with a mean 
deviation of $\sim$1.5 for each Hubble type. In order to test this possibility, we selected a subsample composed by the 142 common objects between AMIGA and the \citet{2010ApJS..186..427N} sample, and calculated the median values of (g$-$r) obtained using each morphological classification. We found that the difference between both median colors in the common sample was consistent with that found using the full samples.

 \begin{figure*}[t]
\centering
      \includegraphics[angle=0,width=17.5cm]{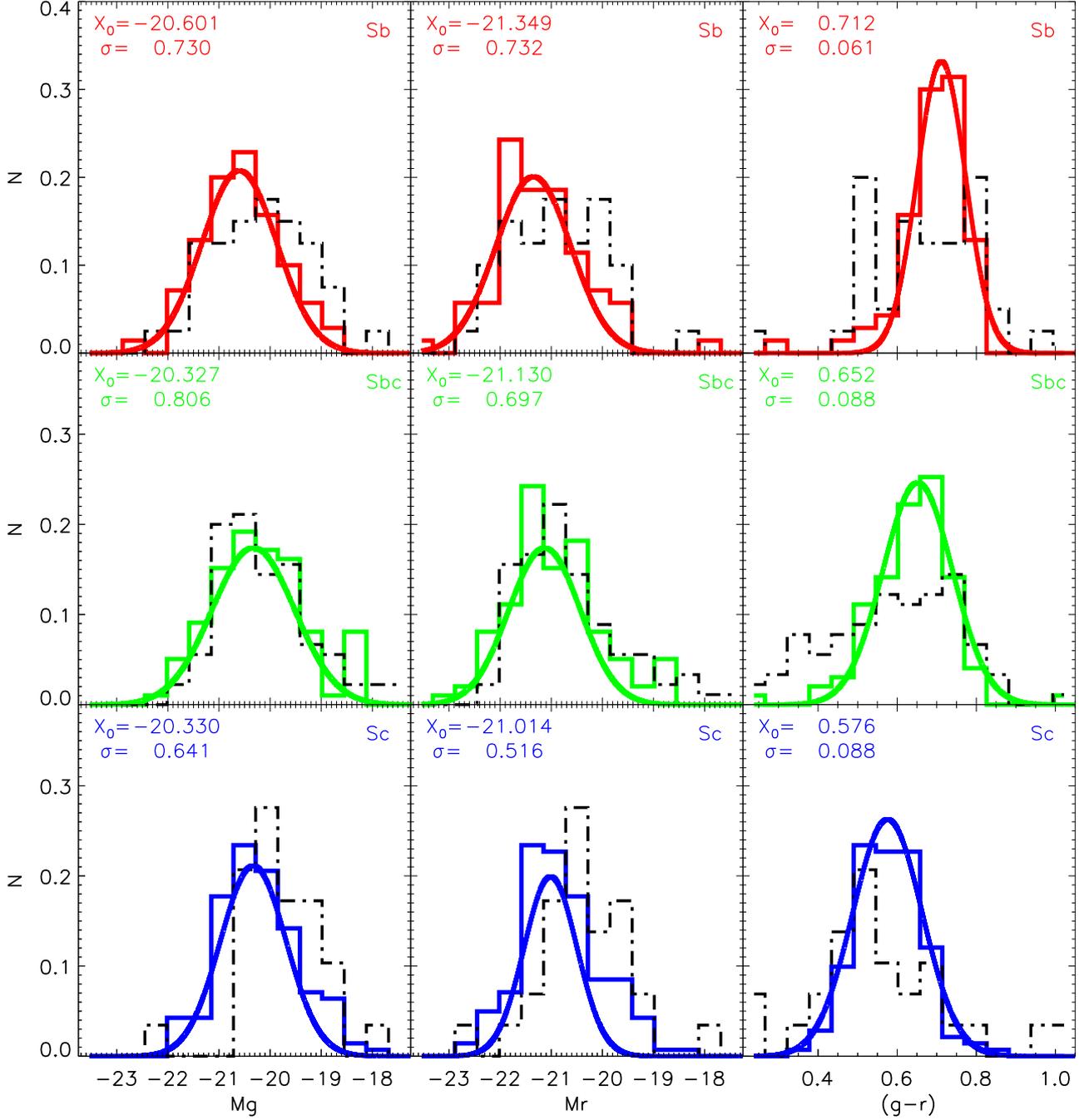}
      \caption{Distribution of absolute magnitudes in the g--band (left), r--band (centre) and (g$-$r) color (right) for all Sb (top), Sbc (middle) and Sc (bottom) galaxies. The solid lines represent the AMIGA--SDSS sample and the dashed lines are the distribution of the CPG close pairs.}
\label{multi_hist}
   \end{figure*}

Nevertheless, the differences in color of close pairs seem to be more robust because the colors of wide pairs are as red as the colors of isolated galaxies. The scatter is also larger for the close pairs, in agreement with \citet{1978ApJ...219...46L}. In Fig.~\ref{multi_hist} we present the distribution of absolute magnitudes in r and g--bands as well as (g$-$r) color for three morphological types (Sb, Sbc and Sc) of the AMIGA--SDSS sample. We have also represented the same distributions for the close pairs of the CPG sample. While the color distribution of each morphological type has gaussian distribution for our sample, the color histogram for the CPG sample follows a non gaussian distribution. The loss of Gaussianity in the distribution of color can be interpreted as an effect of the environmental nurture that occurs in a sample of pairs. 

We have also investigated the change in the median colors when we rejected objects with asymmetries (IA=1) and the differences are negligible.

\section{Conclusions}

We have taken a first look at the colors of isolated galaxies using SDSS g and r magnitudes. A first look is also inevitably an introduction into the pitfalls of using the SDSS automated measures of magnitudes. We find that the color distributions of morphological subtypes can be well described as Gaussian distributions with FWHM (g$-$r)=0.1--0.2. This is expected for samples where effects of environmental nurture have been minimized, and is supported by the fact that this Gaussianity was not observed in the sample of galaxy pairs. The majority of the color dispersion for spiral subtypes is caused by a color--luminosity correlation, with more massive spirals showing redder colors. In fact most of the Sb spirals in our sample concentrate at the upper region of the blue cloud, below the red sequence. We see little evidence for a green valley in our sample with most spirals redder than (g$-$r)=0.7 having spurious colors ($\sim$80$\%$). We find a preliminary difference in the color of the isolated and paired spiral galaxies. The median value of (g$-$r) seems to be bluer when interactions come into play. Nevertheless, no difference is found in the (g$-$r) color of early--type galaxies, which suggests that the bluer color of spirals in pairs is presumably due to interaction enhanced star formation. Our sample of isolated galaxies gives a median absolute deviation in color which is lower than that in pairs of galaxies, where a more active star formation and, perhaps, a higher dust diffusion caused by the interaction, are also sources of color dispersion. The mean colors and dispersions for isolated galaxy subtypes (especially E/S0 and Sb--Sc) are likely the best nurture--free measures so far obtained.

\begin{acknowledgements}

This work has been supported by Grant AYA2008-06181-C02 co-financed by MICINN and FEDER funds, and the Junta de Andalucía (Spain) grants P08-FQM-4205 and TIC-114. We are grateful to the AMIGA team for the great work they have done on improving the sample and to the anonymous referee for useful comments.

Wf4Ever is funded by the Seventh Framework Programme of the European Commission (Digital Libraries and Digital Preservation area ICT-2009.4.1 project reference 270192). We are grateful to all our collaborators in this
project.

We thank the SDSS group for making their catalogues and data publicly available. Funding for SDSS-III has been provided by the Alfred P. Sloan Foundation, the Participating Institutions, the National Science Foundation, and the U.S. Department of Energy. The SDSS-III web site is http://www.sdss3.org/.

SDSS-III is managed by the Astrophysical Research Consortium for the Participating Institutions of the SDSS-III Collaboration including the University of Arizona, the Brazilian Participation Group, Brookhaven National Laboratory, University of Cambridge, University of Florida, the French Participation Group, the German Participation Group, the Instituto de Astrofisica de Canarias, the Michigan State/Notre Dame/JINA Participation Group, Johns Hopkins University, Lawrence Berkeley National Laboratory, Max Planck Institute for Astrophysics, New Mexico State University, New York University, Ohio State University, Pennsylvania State University, University of Portsmouth, Princeton University, the Spanish Participation Group, University of Tokyo, University of Utah, Vanderbilt University, University of Virginia, University of Washington, and Yale University.
 
We thank the SAO/NASA Astrophysics Data System (ADS) that is always so useful. 

 \end{acknowledgements}

\end{document}